\documentclass{aa}

\usepackage{subcaption}
\usepackage{color,xcolor}
\usepackage{float} 

\usepackage{comment}
\usepackage{color}

\usepackage{txfonts}
\usepackage{hyperref}
\hypersetup{colorlinks,linkcolor={blue},citecolor={blue},urlcolor={blue}} 

\usepackage{array}

\begin{document}

\title{Tracing the high-$z$ cosmic web with \emph{Quaia}: Catalogues of voids and clusters in the quasar distribution}

\author{N.~Arsenov\inst{1}\fnmsep\inst{2} \and A.~Kov\'acs\inst{2}\fnmsep\inst{3} \and M.~Pérez Sar\inst{4}\fnmsep\inst{5} \and Á.Sz.~Bogdán\inst{2}\fnmsep\inst{3}\fnmsep\inst{6} \and F.~Sinigaglia\inst{7}\fnmsep\inst{8}\fnmsep\inst{4}\fnmsep\inst{5} \and F.S.~Kitaura\inst{4}\fnmsep\inst{5} \and G.~Favole\inst{4}\fnmsep\inst{5} \and L.~Slavcheva-Mihova\inst{1}}

\institute{
Institute of Astronomy and NAO, Bulgarian Academy of Sciences, 72 Tsarigradsko Chaussee Blvd., 1784 Sofia, Bulgaria\\
\email{\textcolor{blue}{narsenov@nao-rozhen.org}}
\and 
MTA--CSFK \emph{Lend\"ulet} ``Momentum'' Large-Scale Structure (LSS) Research Group, Konkoly Thege Mikl\'os \'ut 15-17, H-1121 Budapest, Hungary
\and 
Konkoly Observatory, HUN-REN Research Centre for Astronomy and Earth Sciences, Konkoly Thege Mikl\'os {\'u}t 15-17, H-1121 Budapest, Hungary
\and Instituto de Astrof\'isica de Canarias, Calle via L\'actea s/n, E-38205, La  Laguna, Tenerife, Spain
\and Departamento  de  Astrof\'isica, Universidad de La Laguna,  E-38206, La Laguna, Tenerife, Spain
\and Institute of Physics and Astronomy, ELTE E\"otv\"os Lor\'and University, P\'azm\'any P\'eter s\'et\'any 1/A
H-1117 Budapest, Hungary
\and Département d’Astronomie, Université de Genève, Chemin Pegasi 51, CH-1290 Versoix, Switzerland
\and Institut für Astrophysik, Universität Zürich, Winterthurerstrasse 190, CH-8057 Zürich, Switzerland
}

\titlerunning{\emph{Quaia} voids and clusters}
\authorrunning{N. Arsenov et al.}

\date{Received July 2025}

\abstract 
{Understanding the formation and evolution of the cosmic web of galaxies is a fundamental goal of both theoretical and observational cosmology, which use various tracers of the cosmic large-scale structure at an ever wider range of redshifts.}
{Our principal aim is to advance the mapping of the cosmic web at high redshifts using observational and synthetic catalogues of quasars , which offer a powerful probe of structure formation and the validity of the concordance cosmological model at the largest scales in the Universe.}
{In this analysis, we selected 708,483 quasars at $0.8<z<2.2$ from the \emph{Quaia} dataset; this enabled an extended reconstruction of the matter density field using 24,372 $deg^2$ sky area with a well-understood selection function, thus going beyond the capacity of previous studies. Using the \texttt{REVOLVER} method, we created catalogues of voids and clusters based on the estimation of the local density at quasar positions with Voronoi tessellation. We tested the consistency of \emph{Quaia} data and 50 realistic mock catalogues, including various parameters of the voids and clusters in characteristic subsets of the data, and also measurements of the density profiles of these cosmic super-structures at $R\approx100~h^{-1}{\rm Mpc}$ scales.}
{We identified 12,820 voids and 41,154 clusters in the distribution of \emph{Quaia} quasars. We found an $\sim5-10\%$ level of agreement between data and the ensemble of the 50 mocks considering void and cluster radii, average inner density, and density profiles at all redshifts. In particular, we tested the role of survey mask proximity effects in void and cluster detection, which, although present in the data, are consistent in simulations and observations. Testing the extremes, the largest voids and clusters reach $R_{\rm eff}\approx250~h^{-1}{\rm Mpc}$ and $R_{\rm eff}\approx150~h^{-1}{\rm Mpc}$, respectively, but without evidence for ultra-large cosmic structures exceeding the dimensions of the largest structures in our mock catalogues.}
{Our data-analysis results highlight the capacity of \emph{Quaia} quasars to robustly map the high-$z$ cosmic web, further supported by the fully consistent statistical results from 50 mock catalogues. As an important deliverable, we share our density field estimation, void catalogues, and cluster catalogues with the public, which allows various additional cross-correlation probes in the high-$z$ cosmic web.}


\maketitle
\footnotetext{This manuscript is published in A\&A: \href{https://doi.org/10.1051/0004-6361/202557361}{doi.org/10.1051/0004-6361/202557361}}

\section{Introduction}
\label{sec:intro}

\begin{figure*}
\centering
\includegraphics[width=95mm]{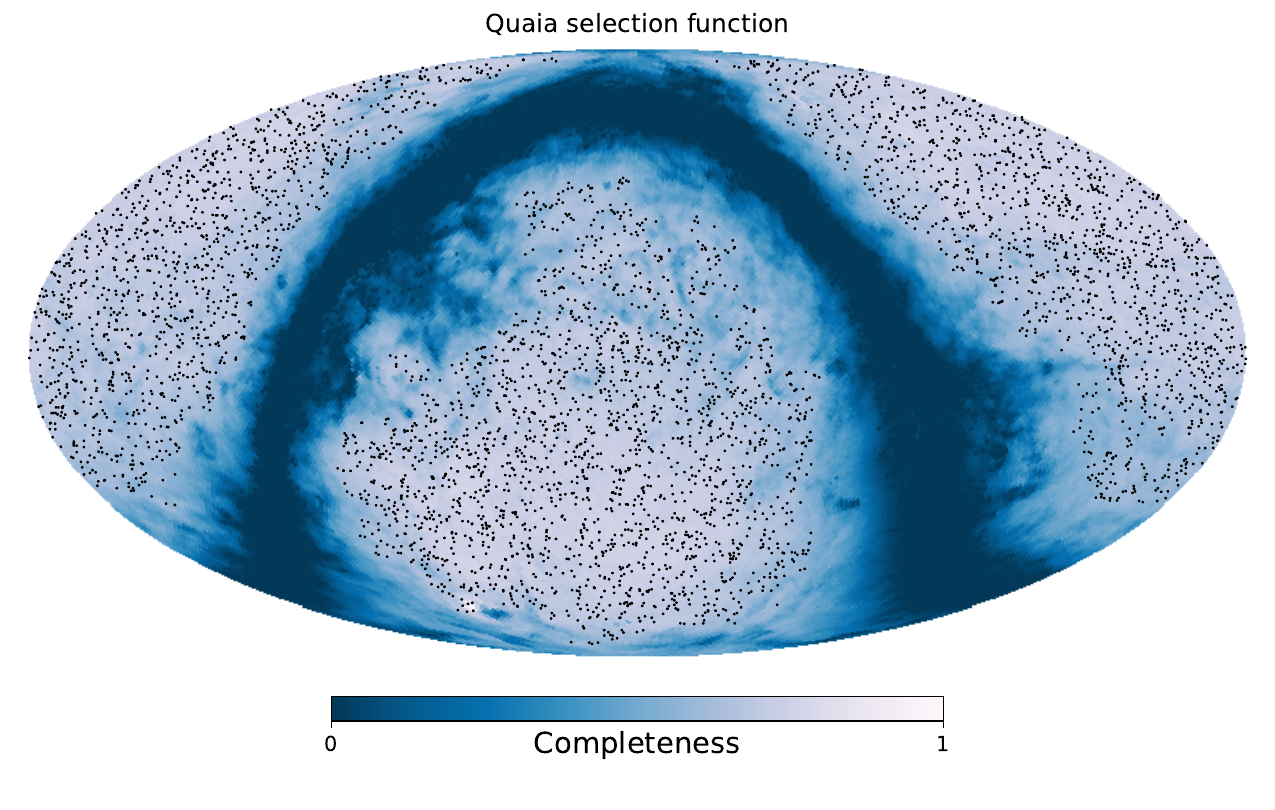} \includegraphics[width=80mm]{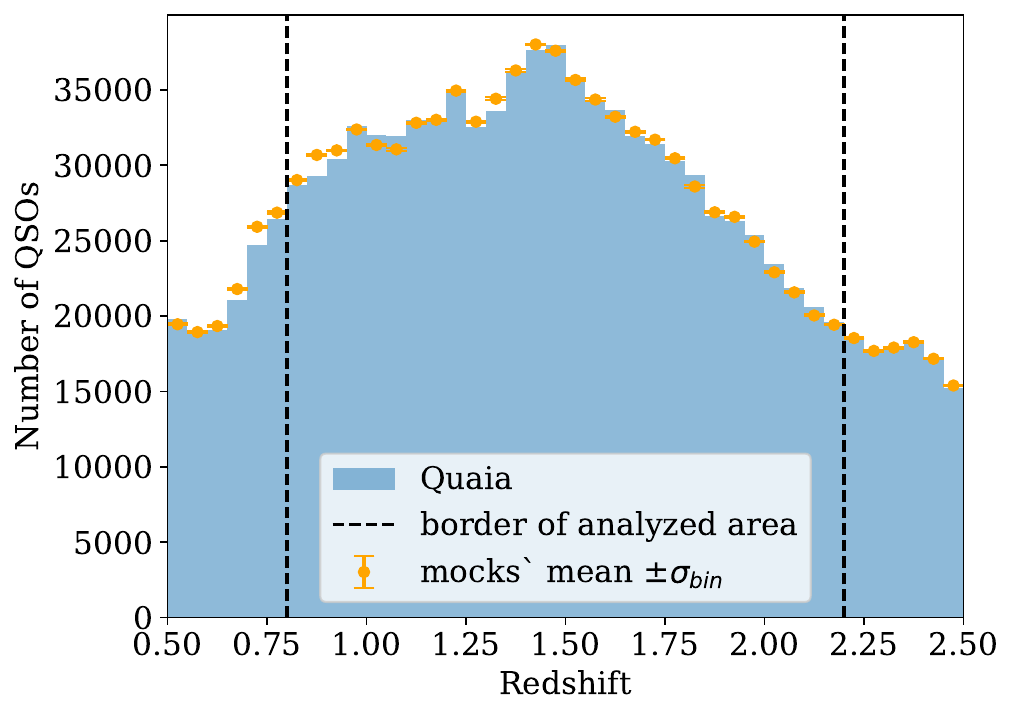}
\caption{Left: \emph{Quaia} selection function. This is the basis of our masking strategy. We also show the distribution of 4520 quasars in a narrow redshift slice at $1.8<z<1.81$ on top. Right: Redshift distribution of \emph{Quaia} quasars, showing the good agreement with mocks.}
\label{fig:selection_function_and_quasars}
\end{figure*}

Mapping out the intricate cosmic web of galaxies is a major goal of observational cosmology that motivates deep and wide sky survey projects \citep[see e.g.][]{DES,DESI,euclid,lsst} as well as various numerical simulation approaches \citep[see e.g.][]{Cai2009,Flagship,Takahashi2017,Racz2023,Schaye2023}. The largest and in many senses most extreme objects in this cosmic large-scale structure (LSS) are the superclusters of galaxies and the cosmic voids, where the density of galaxies is significantly lower than the average. These superstructures appear at 10-100 megaparsec scales and offer insights into gravitational evolution, galaxy formation, and the underlying cosmological model \citep[see e.g.][]{Pisani2019}. Moreover, their spatial distribution and statistical properties also provide a unique opportunity to test the cosmological principle — the assumption that the Universe is statistically homogeneous and isotropic on the largest scales.

In particular, cosmic voids have received increasing attention in recent years as cosmological probes due to their sensitivity to dark energy, modified gravity, neutrino mass, and primordial non-Gaussianity, among other phenomena \citep[see e.g.][]{Clampitt2013,Cai2015,Kitaura2016,Cautun2018,Baker2018,Schuster2019,Davies2021,Contarini2021,Vielzeuf2023}. Voids constrain cosmological models through various probes, such as the void size function, density and velocity profiles, lensing effects, and their evolution with redshift \citep[see e.g.][]{Amendola1999, Krause2013,pisani2015,Sanchez2016,Fang2019,Nadathur2019,Hamaus2021}.

Another active area of research has been to cross-correlate the positions of cosmic voids with the cosmic microwave background (CMB). Their imprints in CMB lensing convergence maps \citep[see e.g.][]{cai2013,Raghunathan2019,Kovacs2022,CamachoCiurana2024,Sartori2024}, in Compton y-maps to study the thermal Sunyaev-Zeldovich (tSZ) effect \citep{alonso18,Li2024}, and in temperature maps via the integrated Sachs-Wolfe (ISW) effect \citep[see e.g.][]{SachsWolfe} have all been measured extensively, often with intriguing tensions \citep[see e.g.][]{Granett2008,Ilic2013,Kovacs2016, NadathurCrittenden2016, Kovacs2019}.

\begin{figure*}
\centering
\includegraphics[width=175mm]{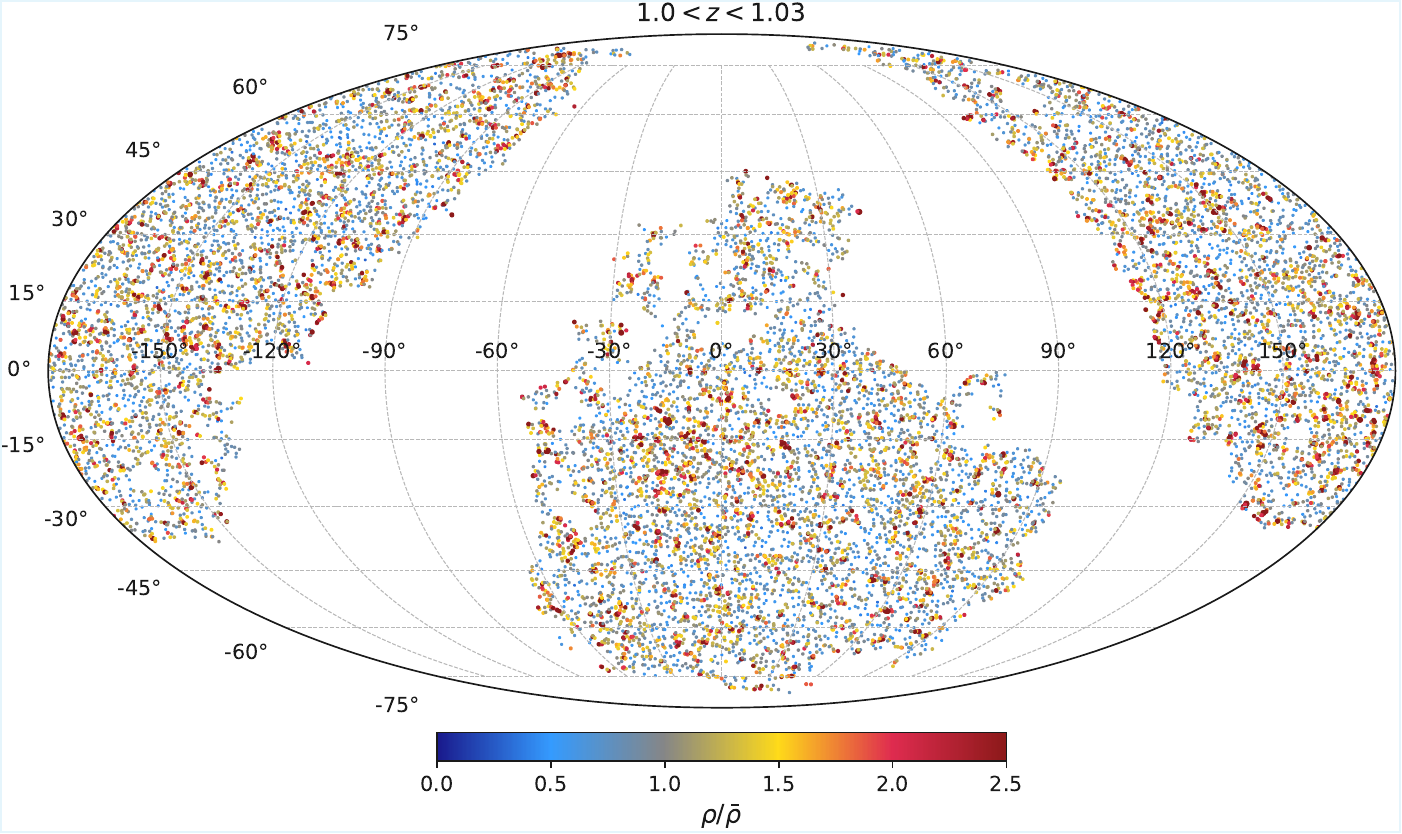}
\caption{Redshift slice of the \emph{Quaia} catalogue at $1.0 < z < 1.03$ in equatorial coordinates. Based on the Voronoi tessellation, the reconstructed local over-density ($\rho/\bar{\rho}$) at the quasar positions is colour-coded and the size of the points is also proportional to their density. Pixels of low completeness are excluded from the analysis using the angular selection function. The map shows large-scale clustering of quasars, without clearly outstanding features.}
\label{fig:quasar_density_in_slice}
\end{figure*}

\begin{figure*}
\centering
\includegraphics[width=155mm]{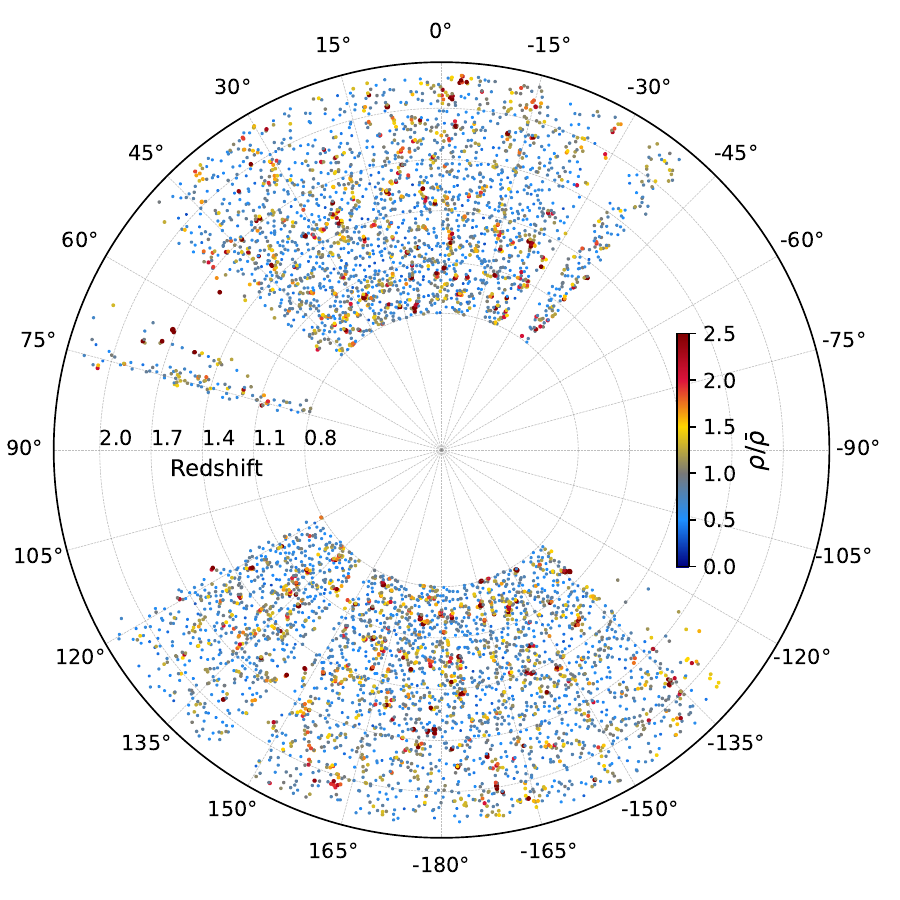}
\caption{Two-dimensional view of the \emph{Quaia} dataset at $0.8 < z < 2.2$ with $-180^{\circ}<RA<180^{\circ}$, but only showing quasars with $-0.5^{\circ}<Dec<0.5^{\circ}$. The over-density ($\rho/\bar{\rho}$) colour-coding and marker sizes are the same as in Fig.~\ref{fig:quasar_density_in_slice}.}
\label{fig:lightcone_map}
\end{figure*}

While low-redshift galaxy surveys such as Sloan Digital Sky Survey
(SDSS), Baryon Oscillation Spectroscopic Survey (BOSS), Dark Energy Survey (DES), and more recently Dark Energy Spectroscopic Instrument (DESI) survey have enabled detailed reconstructions of the galaxy density field and the creation of void catalogues up to $z \sim 0.8$ \citep[see e.g.][]{Mao2017,Douglass2023,Rincon2025}, the intermediate redshift regime $0.8 < z < 2.2 $ in the cosmic web, which is
studied in this analysis,  remains relatively uncharted. In this range, currently available galaxy samples are sparse and quasars provide the best available tracers of the underlying dark matter distribution, due to their brightness. The extended Baryon Oscillation Spectroscopic Survey (eBOSS) facilitated one of the first statistical void analyses at these redshifts using spectroscopic quasars, including the 3D void catalogues by \cite{Hawken2017} and \cite{Aubert2020}, and 2D void catalogues by \cite{Kovacs2021}. While providing a rather small sky coverage for accurate statistics (approx. 4800 $deg^2$ in Data Release 16), these analyses demonstrated the viability of using quasars to probe the cosmic large-scale structure at high redshift, despite their low spatial density.

Recent years have also witnessed growing interest in identifying large-scale overdensities, such as superclusters and quasar groups \citep[see e.g.][]{Park2015, Einasto2021,Liu2024}. Superclusters, the most massive coherent structures in the cosmic web, provide critical insights into the non-linear regime of structure formation and the transition between filamentary and cluster-dominated environments. A number of studies have analysed galaxy and quasar surveys to detect such structures, using methods ranging from friends-of-friends linking algorithms to percolation analysis and minimal spanning trees \citep[see e.g.][]{Libeskind2018,Naidoo2020}. In particular, among several ultra-large structures \citep[see e.g.][]{Lopez2022,Horvath2025}, large quasar groups (LQGs) have been reported at higher redshifts, extending over several hundred comoving megaparsecs \citep[such as the Huge-LQG, see][]{Clowes2014}. While the statistical significance and cosmological implications of these LQGs remain debated, they have raised questions about the validity of the cosmological principle \citep[see e.g.][]{Nadathur2013,Sawala2025} and their detection motivates the use of increasingly large and homogeneous quasar samples for cosmic web analysis.

To advance this line of research, we used the recent \emph{Quaia} quasar catalogue \citep{Storey_Fisher_2024} that provides an all-sky sample of quasars with precise photometric redshifts \citep[see e.g.][for previous applications]{Piccirilli2024,Veronesi2025,Fabbian2025,Alonso2025}. We mapped the cosmic large-scale structure at $0.8 < z < 2.2$ using this catalogue and focused on the identification of the largest voids and (super)clusters traced by the quasar distribution. Since the exact properties of such extreme structures may challenge our understanding of cosmic variance and structure formation in the standard $\Lambda$-cold dark matter ($\Lambda$CDM) model, a more inclusive census of these cosmic superstructures at high redshift from the \emph{Quaia} dataset might help resolve any related tension. We note that extensions of the \emph{Quaia} cosmic web analyses to higher redshifts are also feasible, especially using the next \emph{Gaia} data release (DR4) to possibly increase the quasar source density.

Our analysis results in two main products. First, we estimate the local overdensity at the position of each quasar. Second, we provide a catalogue of voids and clusters identified in the \emph{Quaia} quasar distribution. The combination of these enables future studies to examine quasar properties,
such as luminosity or spectral features, as a function of their cosmic environment. 

The paper is organized as follows. In Section \ref{sec:data}, we introduce our observational and mock datasets, as well as our methodology. Section \ref{sec:results} contains a description of our analysis of the reconstructed density field of the quasars, including catalogues of voids and clusters, followed by a summary of our main conclusions in Section \ref{sec:summary}.

\section{Datasets and methodology}
\label{sec:data}

\subsection{Quasar catalogue}
\label{subsec:quasar_catalogue}
In our cosmographical analysis, we used the \emph{Quaia} quasar catalogue \citep{Storey_Fisher_2024}, based on a cross-match between the \emph{Gaia} Data Release 3 quasar candidates \citep{2023A&A...674A..41G,2023A&A...674A...1G, 2023A&A...674A..31D} and the \emph{Wide-field Infrared Survey Explorer} \citep[\emph{WISE},][]{2014AJ....147..108L} dataset. This cleaned catalogue exploits segmentation in colour-colour spaces to differentiate stars, galaxies, and quasars, and excludes sources with high proper motions. The \emph{Quaia} catalogue comes in two versions; the full dataset contains approximately 1.3 million quasars with $G < 20.5$ (this is what we work with), while a higher fidelity $G < 20.0$ subsample is also characterized with about 760,000 quasars.

Furthermore, we make use of the selection function provided by the \emph{Quaia} team (shown in Fig.~\ref {fig:selection_function_and_quasars}), which is a full-sky \texttt{healpix} \citep{gorski2005healpix} map that assigns the completeness of quasar observations in a given pixel. Completeness is estimated from the most important systematic effects, such as dust extinction, stellar density, and scan patterns of the parent surveys (see \cite{Storey_Fisher_2024} for details). This selection function sky map ($\texttt{sel\_func}$, see also Table~\ref{tab:value_added_table} for further detail) facilitates the necessary corrections to the local density of quasars in noisier pixels when looking for voids and clusters (see Section~\ref{sec:revolver_methodology}), and it also allows us to completely exclude pixels from the analysis. We decided to set a threshold of $\texttt{sel\_func}>0.52$ in the completeness map, which is a conservative choice to keep about 24,372 $deg^2$ of the sky area while excluding the noisiest pixels near the Milky Way's plane.

In the above observational window on the sky, we also applied a redshift cut at $0.8 < z < 2.2$, leaving 708,483 quasars for our analysis, i.e. $55\%$ of the whole quasar dataset. This choice is motivated by the higher number density of sources in the \emph{Quaia} catalogue in that range (see Fig.~\ref {fig:selection_function_and_quasars}), and also to allow more direct comparisons with previous results, since eBOSS quasar analyses of voids also focused on this redshift range \citep[][]{Kovacs2021}.

\subsection{Mapping the quasar density field}
\label{sec:revolver_methodology}
To estimate the local overdensity of quasars ($\rho/\bar{\rho}$, where $\bar{\rho}$ is the mean density) and then identify voids and clusters in their distribution, we used the open-source \texttt{REVOLVER} (REal-space VOid Locations from surVEy Reconstruction) code\footnote{https://github.com/seshnadathur/Revolver} \citep{Nadathur2019}. Based on the \texttt{ZOBOV} watershed algorithm \citep{ZOBOV}, \texttt{REVOLVER} uses a Voronoi tessellation methodology to create a detailed map of the density field (see Fig.~\ref{fig:quasar_density_in_slice} and Fig.~\ref{fig:lightcone_map} for subsets of the resulting quasar overdensity map). The algorithm first assigns a Voronoi volume to each input tracer of the large-scale structure, i.e. marking all points closer to that tracer than to any other (see \texttt{vol\_nocorr} in Table~\ref{tab:value_added_table}). Then, we used the optional tools in \texttt{REVOLVER} to apply weights for individual structures, taking into account survey completeness in pixels or in the redshift distribution (assigning a corrected volume, \texttt{vol\_corr}), in order to correct for known imperfections in the input data.

\begin{figure*}
\centering
\includegraphics[width=180mm]{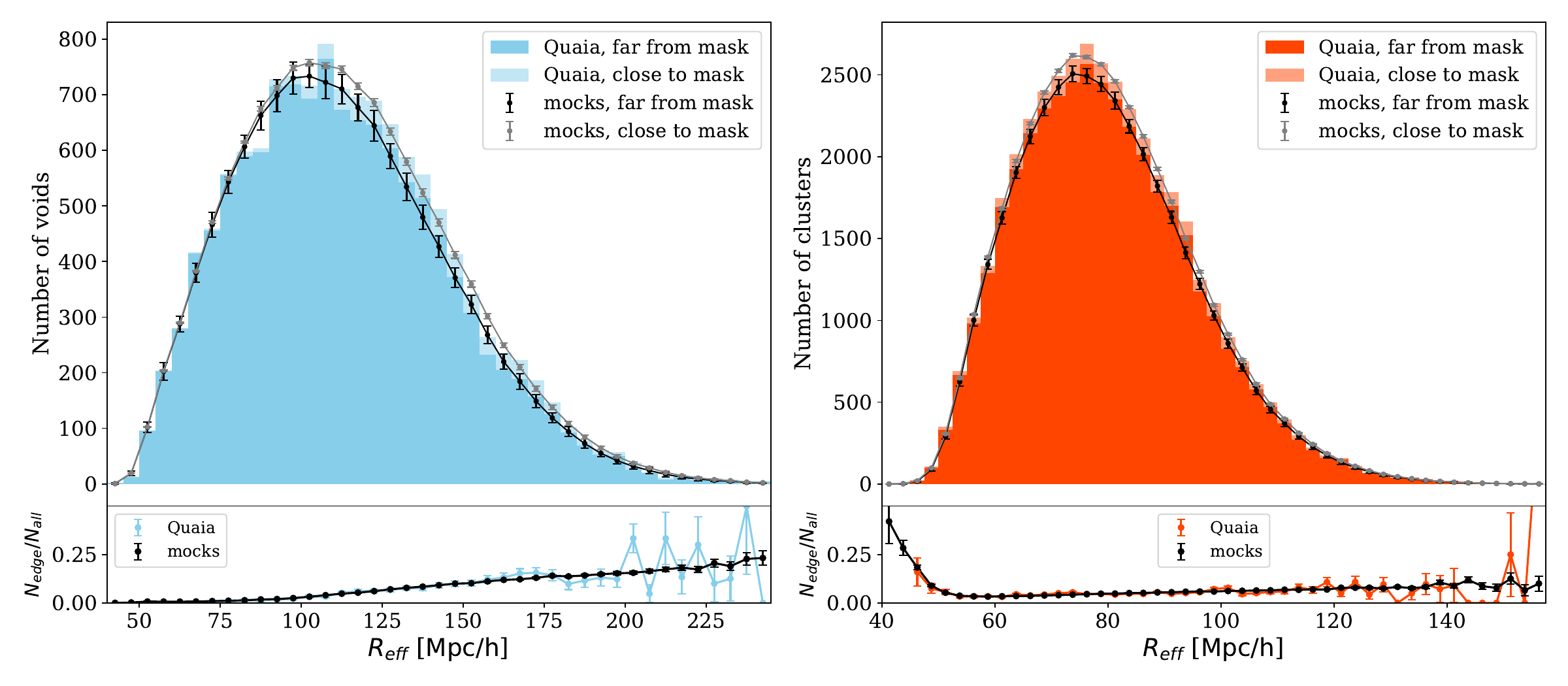}
\caption{Void radii (left) and cluster radii distribution (right) in the \emph{Quaia} catalogue. With full colours, we show structures that are far from the mask ($\texttt{EdgeFlag}=0$), while pale bars on top show the number of voids and clusters close to the survey edge ($\texttt{EdgeFlag}=1$). We found good agreement when comparing the observations with the mean and standard deviation of the mocks (black and grey data points). On the main panels, error bars correspond to the standard deviations of the 50 mock realizations. In the bottom panels, error bars for the mocks are again estimated from the 50 realisations, while for \emph{Quaia} we used the binomial sample standard deviation $\sigma = \sqrt{p(1-p)/N}$ with $p=N_{\rm edge}/N_{\rm all}$ where $N$ is the number of voids in the bin. We also used these error estimations in Figs.~\ref{fig:delta_min}-\ref{fig:redshift}.}
\label{fig:radius}
\end{figure*}

In the next step, the \texttt{ZOBOV} algorithm creates catalogues of voids and clusters from the Voronoi cells by applying the watershed approach. This is analogous to flooding a topographic land surface, where valleys get filled when water levels are rising. Then, neighbouring valleys might merge into larger basins when the water level rises further \citep[see][for more details]{ZOBOV}. To each void and cluster, multiple features are presented as output and we listed the most relevant ones for our interests in Tables~\ref{tab:void_or_cluster_cat} and~\ref{tab:value_added_table}.

Here we note that we considered a flat $\Lambda$CDM cosmology using \texttt{astropy}\footnote{\url{https://www.astropy.org/}} throughout this work, including calculations of distances in the void and cluster finding processes, based on \citet{Planck2018} parameter values: $H_0 = 67.6\,\mathrm{km\,s^{-1}\,Mpc^{-1}}$, $\Omega_{\rm m}= 0.31,$ and $\Omega_{\Lambda} = 0.69$.

When working with \texttt{REVOLVER}, we pruned the \emph{Quaia} input data in the following ways:
\begin{itemize}
    \item We used a binary sky mask to exclude pixels where completeness is low, mostly close to the Galactic plane. As mentioned in Sect. \ref{subsec:quasar_catalogue}, we only kept pixels with $\texttt{sel\_func}>0.52$ completeness, leaving 24,372 $deg^2$ of sky area. The distribution of quasars within this reliable area is depicted in Fig.~\ref{fig:selection_function_and_quasars} and Fig.~\ref{fig:quasar_density_in_slice}. 
    \item In the rest of the sky, we used the \emph{Quaia} selection function to correct the local density estimation, considering various sources of systematic effects \citep[see][for details]{Storey_Fisher_2024}. Effectively, we increased source density by changing the Voronoi volumes of cells based on the completeness information. 
    \item We applied a correction based on the changing $N(z)$ redshift distribution of sources, caused by the expected sensitivity limitation in observing quasars at higher redshifts. Technically, this step also corrects the density field reconstruction by modifying the Voronoi cell volumes based on quasar redshifts and the data sparsity at that redshift.
\end{itemize}

\subsection{Mock catalogues}
In this work, we use $50$ mock catalogues, specifically tailored to reproduce {\it Quaia} observational data. In what follows, we briefly summarize the main features of the mocks and refer to \citet[][]{Sinigaglia_mocks} for the details.

The mock catalogues used in our analysis were generated by adopting the following procedure:
\begin{enumerate}
    \item Generation of full-sky lightcone dark matter field with smooth redshift evolution at $0<z\lesssim 4$: these were obtained through the \texttt{WebOn} code \citep{Sinigaglia_mocks}, implementing the \texttt{ALPT} structure formation model \citep[][]{Kitaura2013} directly on the lightcone; 
    \item Application of a non-linear, non-local, stochastic parametric \texttt{Hicobian} bias model \citep[see,][and references therein]{ColomaNadal2024} to the dark matter fields to generate quasar number counts in cells. This model was first calibrated to fit the clustering of quasar halo occupation distribution mocks, reproducing the 3D clustering measurements from the DESI One-percent Survey \citep[][]{Yuan2024}. Subsequently, the bias was fine-tuned to reproduce the angular clustering measured from {\it Quaia} data \citep[][]{Storey_Fisher_2024}. As a result, mock averages and observed clustering were in excellent, $1\sigma$ agreement level in most angular bins \cite[see][for more detail]{Sinigaglia_mocks};
    \item Application of a suited sub-grid model to assign positions and velocities to the objects: we adopt a simplified version of the sub-grid model described in \cite{ForeroSanchez2024}. Specifically, we assign quasar positions in correspondence with existing dark matter particles and generate the remaining ones by means of a random uniform sampling within the cell. The velocities are modelled as in \cite{Kitaura2012a}, \cite{Kitaura2012b}, \cite{Kitaura2014}, \cite{Kitaura2016}, \cite{Bos2019}, \cite{Sinigaglia2022}, \cite{Sinigaglia2024a}, and \cite{Sinigaglia2024b}, i.e. as the sum of a large-scale coherent flow component --- consisting of the \texttt{ALPT} velocities --- and a small-scale quasi-virialized motion accounting for the fingers-of-God. 
    \item Injection of {\it Quaia} observational systematics: at the last step, we assign  a realistic spectrophotometric error to every quasar, sampled from the distribution measured directly from the data, and then apply both the angular and the radial selection functions, as presented in \cite{Storey_Fisher_2024}.   
\end{enumerate}

In this way, we obtained mock catalogues that closely resemble the main summary statistics from the {\it Quaia} catalogues. In particular, the angular clustering from the mock catalogues was shown to correctly reproduce the one from the data both in configuration and in Fourier space. 

\section{Results and deliverables}
\label{sec:results}

Here we present our main findings about the cosmic web traced by quasars at $0.8<z<2.2$, and we introduce our data products, which we make publicly available for the community, making use of the cosmographical information from this analysis. Our main deliverables are the following:
\begin{itemize}
    \item An estimation of the local overdensity ($\rho/\bar{\rho}$) at quasar positions in the \emph{Quaia} catalogue and its mocks, based on Voronoi tessellation.
    \item The construction of void and cluster catalogues with \texttt{REVOLVER}, both for \emph{Quaia} and its 50 mock catalogues. We note that these clusters are not expected to be virialized overdensities like galaxy clusters but instead extended groups of quasars, possibly in superclusters. We decided to refer to them as clusters to follow the generic notation of \texttt{REVOLVER}.
    \item A value-added catalogue that contains the quasar properties from \emph{Quaia}, plus the above information about the quasar's local density and its relative position within voids or clusters.
\end{itemize} 

\begin{figure*}
\centering
\includegraphics[width=180mm]{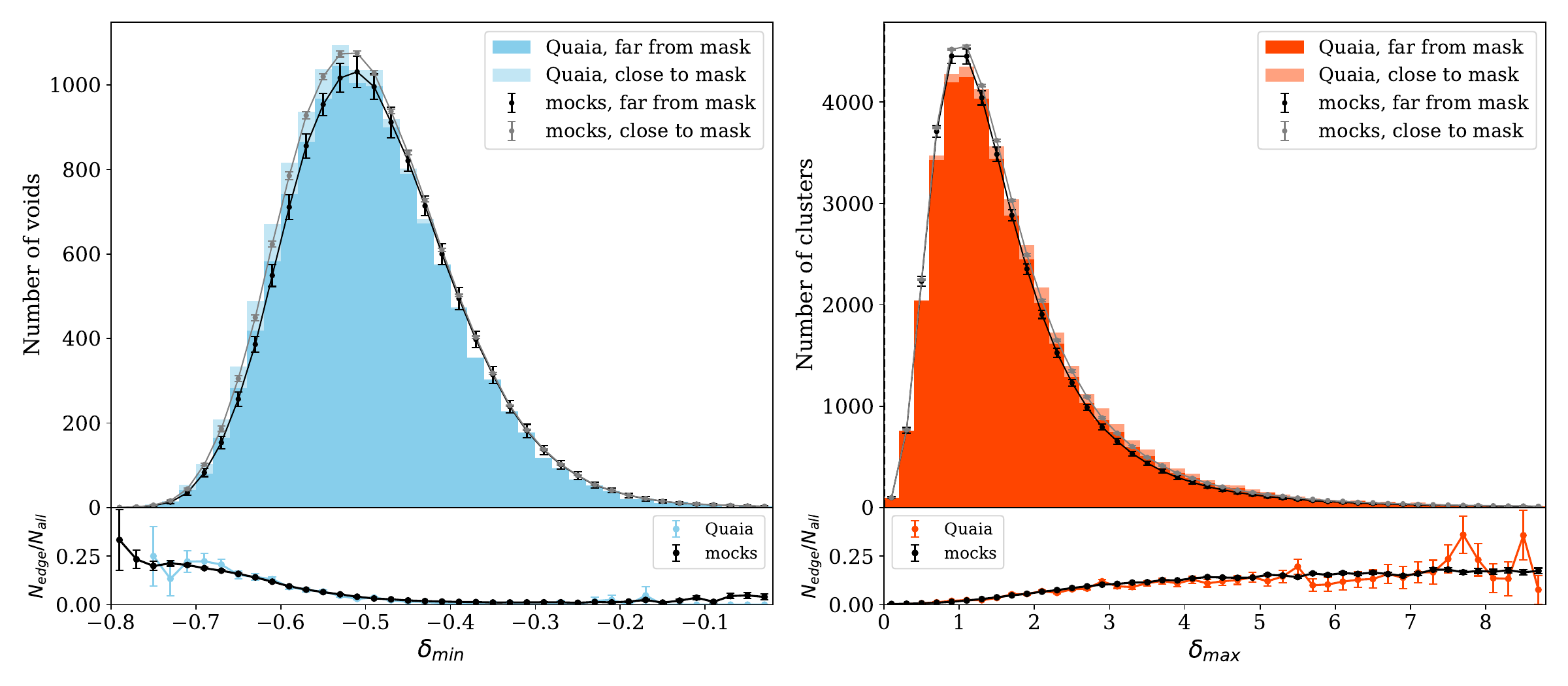}
\caption{Distributions of minimum density in void centres (left) and maximum density in cluster centres (right) in the \emph{Quaia} catalogue. We again compare structures near and far from the survey edges, and also assess consistency between data and mocks.}
\label{fig:delta_min}
\end{figure*}

\begin{figure*}
\centering
\includegraphics[width=180mm]{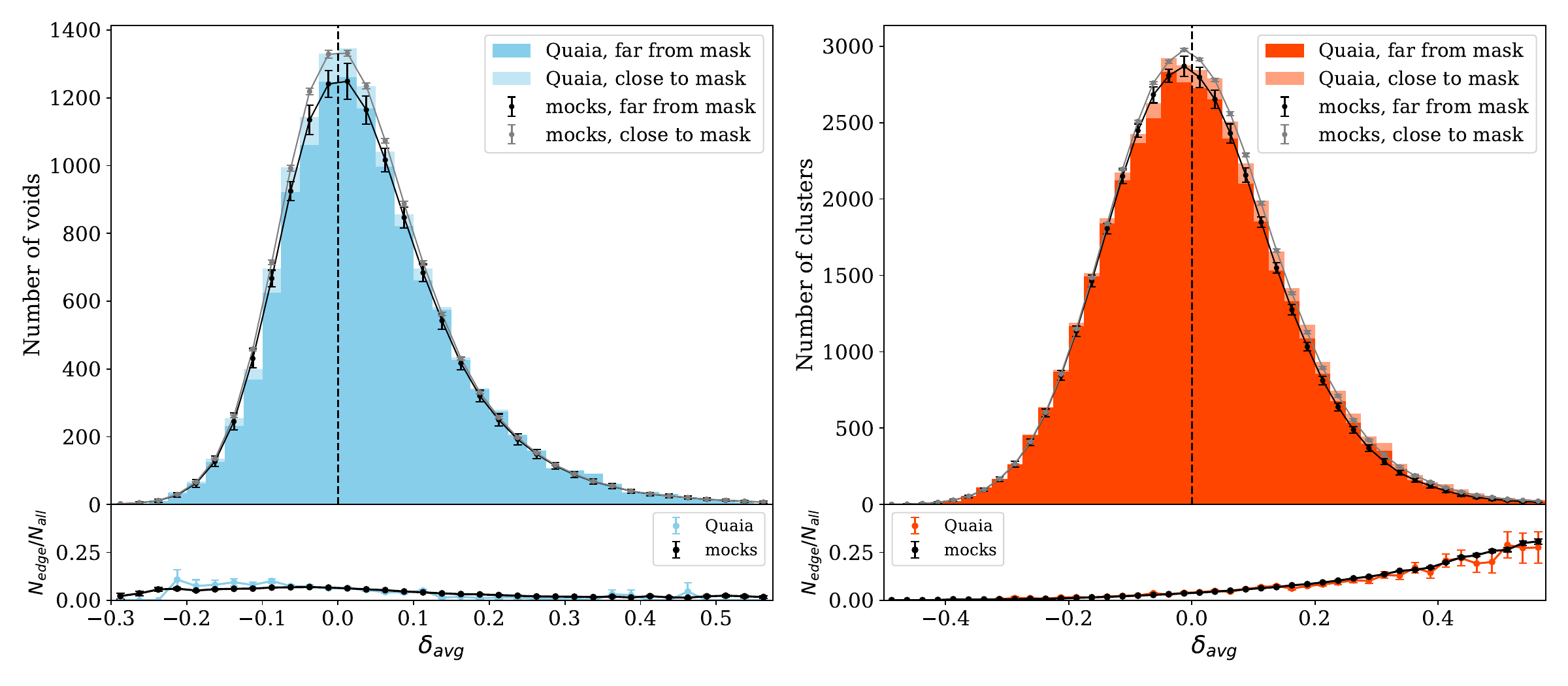}
\caption{Distributions of average density in voids (left) and average density in clusters (right) in the \emph{Quaia} catalogue. We again compare structures near and far from the survey edges, and also assess consistency between data and mocks.}
\label{fig:delta_avg}
\end{figure*}

\subsection{Voids and clusters in the quasar distribution}

As outlined in Section \ref{sec:data}, the fundamental step in our cosmographical analysis of the \emph{Quaia} catalogue was the estimation of the local density, based on the \texttt{ZOBOV} algorithm (see Fig.~\ref{fig:quasar_density_in_slice} and Fig.~\ref{fig:lightcone_map} for subsets of the density reconstruction). Then, we defined extended coherent patterns in the quasar density field as voids and clusters, and we created catalogues of such large-scale structures. 

The format of these catalogues of voids and clusters in the quasar distribution is described in Table~\ref{tab:void_or_cluster_cat}. Columns include positions (RA, Dec, z), effective radius ($R_{\rm eff}$), various density parameters (expressed as $\delta_{\rm min}$ minima and $\delta_{\rm avg}$ average of density fluctuation, with $\delta=\rho/\bar{\rho}-1$), and also a binary \texttt{EdgeFlag} parameter to determine if the given void or cluster is close to the edge of the survey mask. Investigating the $\texttt{EdgeFlag}$ effects takes a significant part of this analysis because voids and clusters might show spurious signals where the data quality is lower near the survey mask. We note that future users of these catalogues might prune them based on their own preferences and explore the survey boundary effects. For further details about these void and cluster parameters see for example \cite{ZOBOV} and \cite{Nadathur2016}.

\begin{figure*}
\centering
\includegraphics[width=180mm]{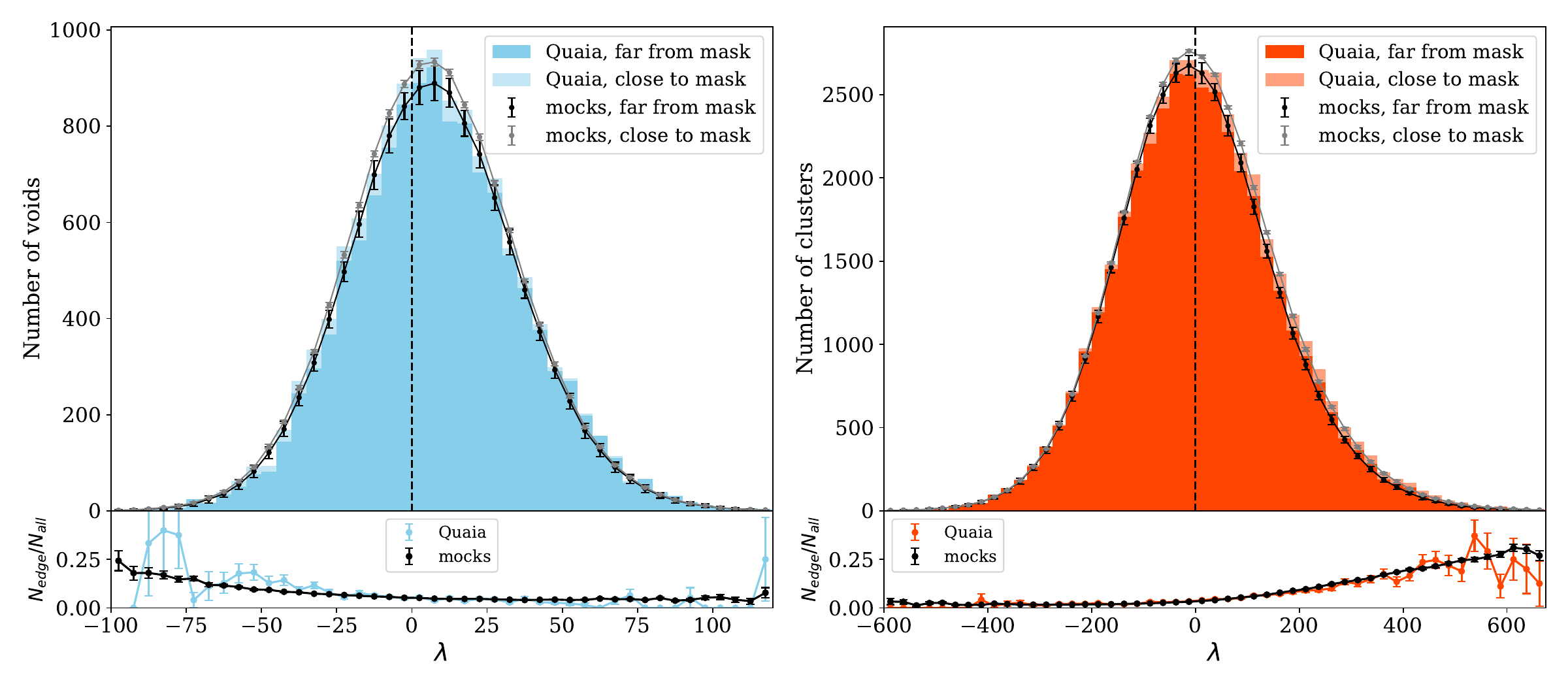}
\caption{Distributions of the $\lambda_v$ parameter values in voids (left) and the $\lambda_c$ parameter values in clusters (right) in the \emph{Quaia} catalogue. We again compare structures near and far from the survey edges, and also assess consistency between data and mocks.}
\label{fig:lambda}
\end{figure*}

\begin{figure*}
\centering
\includegraphics[width=180mm]{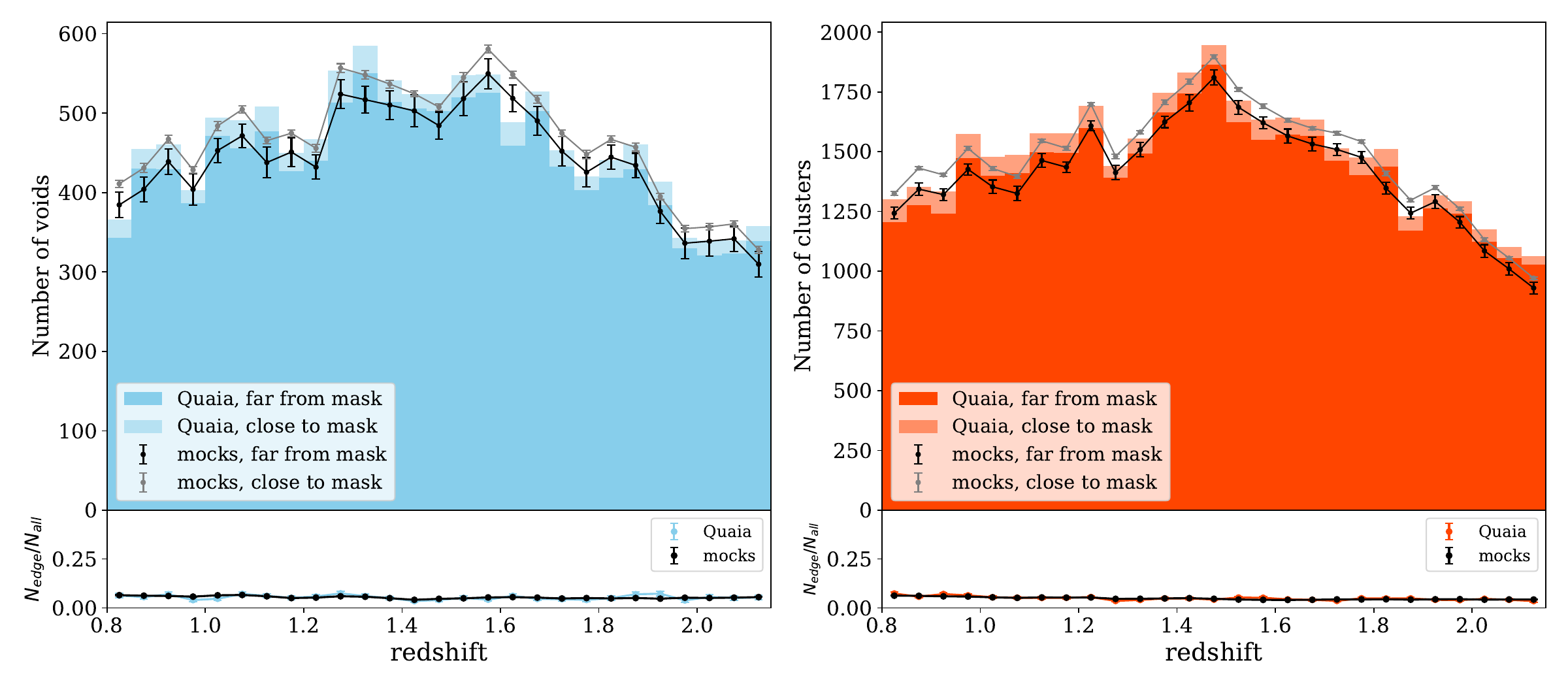}
\caption{Redshift distribution of voids (left) and clusters (right) in the \emph{Quaia} catalogue. We again compare structures near and far from the survey edges, and also assess consistency between data and mocks.}
\label{fig:redshift}
\end{figure*}

In Figs.~\ref{fig:radius}-\ref{fig:redshift}, we present a detailed comparison of the \emph{Quaia} void and cluster catalogue parameters with the 50 mock catalogues that we analysed, given their mean and standard deviation. Each histogram shows an $\sim5-10\%$ level of agreement between observations and simulations, including distributions of their radii, average density, minimum or maximum density, and redshift distribution. We note that we constructed the histogram bars in such a way that voids and clusters with $\texttt{EdgeFlag}=1$ appear on the top of the $\texttt{EdgeFlag}=0$ part of each bar, showing their contribution to the overall count (similarly for mocks, using error bars). Here we list and explain a few characteristic features of our catalogues:

\begin{itemize}
    \item In total, we identified 12,820 voids and 41,154 clusters in the distribution of 708,483 quasars in \emph{Quaia}, which is fully consistent with the typical yield from the mock catalogues.
    \item On average, clusters are more compact than voids and their central density fluctuation is also higher (see Figs.~\ref{fig:radius} and \ref{fig:delta_min}).
    \item Both voids and clusters are considered spherical on average but individually they might have highly irregular shapes.
    \item A typical distinction between different classes of voids is voids-in-voids versus voids-in-clouds, depending on their large-scale environment. Parameters like $\delta_{\rm avg}$ and $\lambda$ provide proxies \citep[see e.g.][]{Raghunathan2019} for such a classification for voids, and for clusters as well (see Figs.~\ref{fig:delta_avg} and \ref{fig:lambda}).
\end{itemize}

Furthermore, we studied the level of \emph{Quaia} versus mock agreement for voids and clusters located near the survey edge ($\texttt{EdgeFlag}=1$), which might contaminate the sample due to their imperfect mapping (see Figs.~\ref{fig:radius}-\ref{fig:redshift}). We found that:
\begin{itemize}
    \item the largest voids and clusters are more prone to edge effects (see tails in the bottom panels of Fig.~\ref{fig:radius}). This is naturally expected, since larger structures have higher chances of touching the survey boundaries.
    \item Voids with very low minimum density ($\delta_{\rm min}\lesssim-0.6$) and clusters with more high maximum density ($\delta_{\rm max}\gtrsim2$) are more sensitive to survey edge effects (see Fig.~\ref{fig:delta_min}, and also Fig.~\ref{fig:delta_avg} for related findings for their mean densities). The enhanced sensitivity of these structures is a result of correlations between their density and radius.
    \item Considering the $\lambda$ parameter, the most extreme voids and clusters show the largest fraction of edge-affected objects (see Fig.~\ref{fig:lambda}). This sensitivity is again due to the fact that $\lambda$ is strongly correlated with the radius (see Table \ref{tab:void_or_cluster_cat}).
    \item While on average there is no significant trend in the redshift distribution of $\texttt{EdgeFlag}=1$ in voids, we found that above $z\approx2$ their ratio slightly rises, most probably due to falling quasar number densities (see Fig.~\ref{fig:redshift}).
    \item Overall, we report good agreement between \emph{Quaia} and the mock datasets in the context of edge effects.
\end{itemize}

We highlight that this excellent agreement was not guaranteed based on the mock construction, which was mostly calibrated on the two-point correlation functions and lower-level cosmic web environment statistics. Therefore, our results further confirm the robustness of the \emph{Quaia} mocks at a different level of complexity in the data analysis.

\begin{figure*}
\centering
\includegraphics[width=180mm]{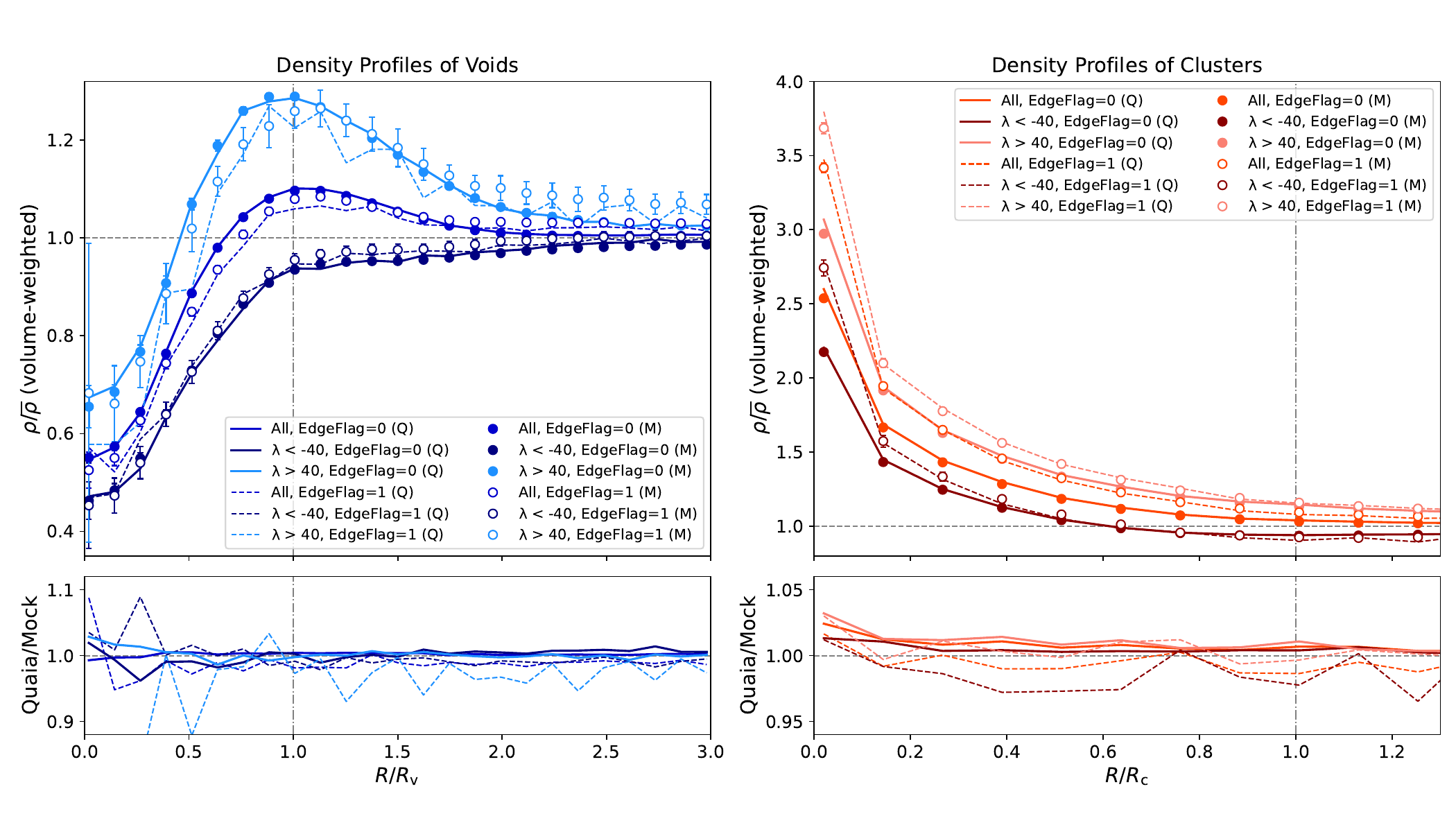}
\caption{Density profiles of voids (left) and clusters (right) in the \emph{Quaia} catalogue. We again compare structures near and far from the survey edges, and also assess consistency between data (Q) and mocks (M) in the bottom panels. We show the density profiles using all voids and clusters, and we also split the catalogues into subsets with extreme values of $\lambda_v$ and $\lambda_c$, as a proxy for their environment.}
\label{fig:density_prof}
\end{figure*}

\subsection{Statistics of the largest voids and clusters}

The typical radius of voids is about $R_{\rm eff}\approx100~h^{-1}{\rm Mpc}$, and $R_{\rm eff}\approx80~h^{-1}{\rm Mpc}$ for clusters. In agreement with the mock statistics, the largest voids reach $R_{\rm eff}\approx250~h^{-1}{\rm Mpc}$ radii, while the largest clusters are about $R_{\rm eff}\approx150~h^{-1}{\rm Mpc}$ in radius. 

As an indicator of the sparsity of the data, the approximate mean particle separation is $d_{\rm mps}=\bar{n}^{-1/3}\approx45~h^{-1}{\rm Mpc}$ at $z\approx0.8$, based on $\bar{n}\approx1.2\cdot10^{-5}~h^{3}{\rm Mpc}^{-3}$ tracer density, which increases to $d_{\rm mps}\approx63~h^{-1}{\rm Mpc}$, based on a $\bar{n}\approx4.3\cdot10^{-6}~h^{3}{\rm Mpc}^{-3}$ source density at $z\approx2.2$. In turn, the usual assumption is that structures below $R_{\rm eff}\approx 2\cdot d_{\rm mps}$ are possibly spurious, and they should be excluded from the subsequent statistical analysis \citep[see e.g.][]{Hamaus2016}. These structures are nevertheless present in our output catalogues, for the sake of completeness, and we leave it to future users to implement the desired pruning cuts that are suitable for their concrete applications.

Considering the high-end tail of the void and cluster radius distribution (see Fig.~\ref{fig:radius}), we again note that the largest structures are most prone to contamination from masking effects, and only the clean $\texttt{EdgeFlag}=0$ subset should be considered for statistical analyses. All things considered, we did not find any evidence of outstanding, ultra-large quasar groups or giant empty voids, neither in the \emph{Quaia} quasar distribution, nor in its 50 mock catalogues. This conclusion was also strengthened by visual inspection using different redshift bins and wedges in the data, in the spirit of Figs.~\ref{fig:quasar_density_in_slice} and \ref{fig:lightcone_map}. We leave the more detailed and formal statistical analysis of the largest structures for future work.

\subsection{A value-added quasar catalogue}

With the intention to create a value-added catalogue of quasars, we combined the information on the local density at quasar positions with the quasar's relative position within voids or clusters. This is relevant because $\rho/\bar{\rho}$ itself is not a unique indicator of void or cluster membership and large-scale environment. We note that there are deeper and shallower voids in the catalogue (based on $\delta_{\rm avg}$ and $\lambda$) where a quasar with a given density might be located near the void's centre or at its outskirts close to its compensation wall \citep[for more information about void types and the role of their environment, see e.g.][]{Raghunathan2019}.

This combined dataset is presented in Table~\ref{tab:value_added_table}, with the following information in three subsections:

\begin{itemize}
    \item First, we listed the main outputs from the tessellation: raw Voronoi volume around the quasar, corrected Voronoi volume (based on selection functions), normalized density ($1/ V_{\rm corr}$), Gaia source ID, unWISE object ID, the selection function value at the position of the quasar, and a quality flag. 
    \item The bottom two sections of Table~\ref{tab:value_added_table} contain information about the host void and/or cluster of the given quasar, listing the most important void and cluster parameters that we also provide in Table~\ref{tab:void_or_cluster_cat} (\texttt{REVOLVER} allows for a quasar to be both a member of a void and a cluster, as these catalogues are constructed in two separate watershed runs on the data). 
    \item Based on additional information on membership in voids and clusters provided by \texttt{REVOLVER}, we calculated the $R/R_{\rm eff}$ \emph{relative} position of each quasar in its host structure, labelled as \texttt{R\_over\_Rv} for voids and \texttt{R\_over\_Rc} for clusters. 
    \item For voids, we provide coordinates both circumcentre (defined by the quasar with the most extreme density and its neighbouring cells near the centre) and barycentre definitions, while for clusters we only provide circumcentres given \texttt{REVOLVER}'s default setting. 
\end{itemize}

The extensive list of columns is intended to help future users make elaborate cuts in the data for their own purposes. One may filter this value-added quasar catalogue by local quasar density, data quality given the selection function at the quasar's pixel, redshift of the quasar, or the radius of the host void or cluster to which the quasar belongs; this can lead to various applications. In particular, the radio loudness of quasars and its dependence on local density might also be studied \citep[see e.g.][]{Arsenov2024} as yet another application of the \emph{Quaia} catalogue.

\subsection{Density profiles of voids and clusters}

To further characterize our void and cluster catalogues, we also measured their quasar number density profiles, again in comparison with the mocks. As we noted above, the \emph{Quaia} catalogue is rather sparse, with $\bar{n}\sim10^{-5}~h^{3}{\rm Mpc}^{-3}$, which limits our capacity to provide a detailed reconstruction of the true underlying matter density field. However, the large number of voids ($N_v=$ 12,820) and clusters ($N_c=$ 41,154) in our sample allows us to provide a precise measurement of the stacked density profile for tens of thousands of cosmic superstructures, which further probes the consistency between simulations and observations.

We decided to use the Voronoi tessellation field estimator (VTFE) method to calculate the density profiles, as opposed to cruder counts-in-shells estimates, which are more prone to Poisson noise \citep[see e.g.][]{Nadathur2014}. This is a typical choice when using void and cluster catalogues detected with the \texttt{ZOBOV} methodology. We used the following volume-weighted estimator for the stacked density in the $j^{\mathrm{th}}$ radial shell from the void or cluster centre, which makes use of the VTFE reconstructed density information:
\begin{equation}
    \label{eq:VTFE1}
\overline{\rho}^j = \frac{\sum_{i=1}^{N_v}\sum_{k=1}^{N_i^j} \rho_k V_k}{\sum_{i=1}^{N_v}\sum_{k=1}^{N_i^j} V_k}\,,
\end{equation}where $V_k$ is the volume of the Voronoi cell of the quasar $k$ and $\rho_k$ is its density inferred from the inverse of the Voronoi volume; the sum over $k$ runs over all quasars in the $j$th shell of void or cluster $i$ (not only void or cluster member quasars); and the sum over $i$ includes all voids or clusters ($N_v$ or $N_c$) in the stack. We used 25 radial bins up to $R/R_v=3$ to measure the shapes of the density profiles in sufficient detail.

Our findings are presented in Fig.~\ref{fig:density_prof}, including a detailed comparison between density profiles measured from \emph{Quaia} versus the mean and standard deviation of the 50 mock catalogues. As in previous figures, we also compared \texttt{EdgeFlag=0} and \texttt{EdgeFlag=1} cases for both voids and clusters. We then further explored the data by splitting the catalogues on the $\lambda$ parameter, isolating subsets of voids-in-voids ($\lambda<-40$) and voids-in-clouds ($\lambda>40$) and similar subsets for clusters in dominantly under-dense and over-dense environments. We confirm the robustness of our mapping of high-$z$ cosmic web using the \emph{Quaia} quasars, as well as using its realistic mock catalogues, by drawing the following conclusions:

\begin{itemize}
    \item In spite of the sparse quasar distribution, we find rather smooth and significantly under-dense profiles for voids, and over-dense profiles for clusters.
    \item The agreement between \emph{Quaia} and the mocks is excellent (approx. $5-10\%$), both for voids and clusters. 
    \item As expected, \texttt{EdgeFlag=1} voids and clusters show more noisy and somewhat distorted profiles compared to the clean \texttt{EdgeFlag=0} subset. However, the agreement between mocks and data remains excellent in this aspect as well.
    \item We found a clear separation between subsets of voids and clusters, selected with different $\lambda$ parameter cuts. 
\end{itemize}

\section{Summary and conclusions}
\label{sec:summary}

In this cosmographical analysis, we took the \emph{Quaia} catalogue of quasars as an input and mapped the cosmic web at redshifts $0.8<z<2.2$. While quasar catalogues only allow a rather sparse sampling of the underlying matter distribution, our motivation was to go beyond the current state-of-the-art in high-$z$ void finding in the quasar distribution \citep[see][for eBOSS DR16 results]{Aubert2020,Kovacs2021}. We thus created a value-added dataset for the 708,483 quasars that we analysed, and our main analysis steps were the following:

\begin{itemize}
    \item Taking into account survey systematics through selection functions (quasar redshift distribution, completeness in pixels), we used the \texttt{REVOLVER} algorithm to estimate the local density at the positions of the quasars (see Figs.~\ref{fig:quasar_density_in_slice} and \ref{fig:lightcone_map}).
    \item  We then built a catalogue of 12,820 voids and 41,154 clusters in the \emph{Quaia} quasar distribution based on a Voronoi tessellation algorithm, using 24,372 $deg^2$  of the sky area.
    \item Importantly, we compared our observational results with 50 mock catalogues, in terms of void and cluster radii, mean, minimum and maximum densities, and redshift distribution, finding an excellent $\sim5-10\%$ level of agreement (see Figs.~\ref{fig:radius}-\ref{fig:redshift}).
    \item For completeness, we estimated the density profiles of the voids and clusters in our catalogue, again comparing observational and synthetic data. For different subsets based on edge effects and void or cluster environments, we again found good agreement (see Fig.~\ref{fig:density_prof}).
\end{itemize}

The final deliverable of our work is a combination of the local density estimation ($\rho/\bar{\rho}$) with the information about the membership of \emph{Quaia} quasars in voids and clusters ($R_{\rm eff}, \delta_{\rm min}, \lambda_{v}$ etc.). This way, it becomes possible to label the quasars based on either of these cosmic web environment parameters, or their combinations, and thus create subsets that are located in over-dense or under-dense environments, even specifying their relative positions within voids ($R/R_v$) or clusters ($R/R_c$). 

We foresee various applications, including cross-correlations with CMB maps, radio catalogues, or data at other wavelengths, which motivates the public release of our value-added catalogues. This dataset will contribute to the full exploitation of the \emph{Quaia} quasar catalogue in even greater detail in terms of cosmic web mapping at high redshift.

\section*{Data availability}
The \emph{Quaia} quasar catalogue and the corresponding selection function map are made publicly available\footnote{\url{https://doi.org/10.5281/zenodo.10403370}} by their authors \citep{Storey_Fisher_2024}. The \texttt{REVOLVER} code is also available publicly\footnote{\url{https://github.com/seshnadathur/Revolver}}, with documentation and examples to run it on synthetic or observational data sets. The main products from this work, presented in Table~\ref{tab:void_or_cluster_cat} and Table~\ref{tab:value_added_table}, are only available in electronic form at the CDS via anonymous ftp to \url{cdsarc.u-strasbg.fr} (130.79.128.5) or via \url{http://cdsweb.u-strasbg.fr/cgi-bin/qcat?J/A+A/}. Code for reproducing our analysis and figures is publicly available at \url{https://doi.org/10.5281/zenodo.16359456}. We are available for consultation about the results or our methodology.

\begin{acknowledgements}
The Large-Scale Structure (LSS) research group at Konkoly Observatory has been supported by a \emph{Lend\"ulet} excellence grant by the Hungarian Academy of Sciences (MTA). This project has received funding from the European Union’s Horizon Europe research and innovation programme under the Marie Skłodowska-Curie grant agreement number 101130774. Funding for this project was also available in part through the Hungarian National Research, Development and Innovation Office (NKFIH, grant OTKA NN147550). L.S.-M was partially supported by the Bulgarian Ministry of Education and Science under Agreement D01-326/04.12.2023. The authors thank Kate Storey-Fisher and the \emph{Quaia} team for their help with the input quasar catalogue.
\end{acknowledgements}

\bibliographystyle{aa}
\bibliography{refs}

\appendix
\section{Data Tables}

\begin{table*}[t]
    \centering
    \caption{Properties of the \emph{Quaia} void and cluster catalogues.}
    \begin{tabular}{>{\raggedright\arraybackslash}p{1.5cm} p{6.0cm} 
                >{\centering\arraybackslash}p{1.2cm} >{\centering\arraybackslash}p{1.2cm} >{\centering\arraybackslash}p{1.3cm} 
                >{\centering\arraybackslash}p{1.3cm} >{\centering\arraybackslash}p{1.2cm} >{\centering\arraybackslash}p{1.2cm}}
        \hline \hline
        {\bf Catalogue} & {\bf Description} & \multicolumn{3}{c}{{\bf Voids}} & \multicolumn{3}{c}{{\bf Clusters}} \\
        {\bf Column} & & Minimum & Median & Maximum & Minimum & Median & Maximum \\ \hline
\texttt{ID} & ID of structure, given by \texttt{ZOBOV} & 116 & 17780.50 & 51259 & 263 & 38794 & 94821 \\ 
\texttt{ra} & right ascension of structure centre [deg] & 0.03 & 172.90 & 359.99 & 0.00 & 172.92 & 360.00 \\ 
\texttt{dec} & declination of structure centre [deg] & -81.69 & 1.21 & 83.84 & -82.72 & 1.06 & 85.09 \\ 
\texttt{redshift} & \emph{Quaia} redshift of structure centre & 0.80 & 1.46 & 2.20 & 0.80 & 1.46 & 2.20 \\ 
\texttt{R\_eff\_mpc} & effective radius of structure [$\mathrm{Mpc/h}$] & 43.11 & 111.48 & 266.91 & 45.06 & 78.34 & 157.26 \\ 
\texttt{delta\_ext} & min. cell density in void, max. in cluster & -0.76 & -0.51 & -0.01 & 0.04 & 1.47 & 49.80 \\ 
\texttt{delta\_avg} & average density ($\delta_{\rm avg}$) within the structure & -0.32 & 0.02 & 0.93 & -0.49 & 0.00 & 1.19 \\ 
\texttt{lambda} & $\delta_{\rm avg} \cdot R^{1.2}$ for voids, $\delta_{\rm avg} \cdot R^{1.6}$ for clusters & -118.72 & 6.78 & 135.89 & -656.00 & 1.27 & 1217.48 \\ 
\texttt{DensRatio} & density ratio of structure & -1 & 1.13 & 2.51 & 1.00 & 1.45 & 25.78 \\ 
\texttt{Theta\_eff} & effective angular size of structure [deg] & 0.89 & 2.19 & 5.52 & 0.81 & 1.54 & 3.64 \\ 
\texttt{EdgeFlag} & 1 if close to the survey edge, else 0 & 0 & 0 & 1 & 0 & 0 & 1 \\ 
\texttt{Nmembers} & number of quasars in structure & 5 & 38 & 379 & 5 & 13 & 78 \\ 

        \hline
    \end{tabular}
\tablefoot{The full tables of 12,820 voids and 41,154 clusters are available at the CDS.}
\label{tab:void_or_cluster_cat}
\end{table*}

\begin{table*}[t]
    \centering
    \caption{Value-added catalogue for all \emph{Quaia} quasars in our analysis}
    \begin{tabular}{>{\raggedright\arraybackslash}p{2.5cm} p{9cm} c c c}
        \hline \hline
        {\bf Catalogue Column} & {\bf Description} & {\bf Minimum} & {\bf Median} & {\bf Maximum} \\ \hline
\texttt{ra\_qso} & right ascension [deg] & 0.00 & 173.37 & 360.00 \\
\texttt{dec\_qso} & declination [deg] & -86.99 & 1.28 & 86.86 \\
\texttt{z\_qso} & \emph{Quaia} redshift & 0.80 & 1.46 & 2.20 \\
\texttt{dens} & normalized density around quasar ($\rho/\overline{\rho} \ , \  \rho = 1/ V_{\rm corr}$) & -1.00 & 0.79 & 40.34 \\
\texttt{vol\_corr} & Voronoi volume corrected for systematics ($V_{\rm corr}$) & -1.00 & 0.87 & 4.10 \\
\texttt{vol\_dens\_flag} & flag if $V_{\rm corr}$ and density are good (1) or bad (0) & 0 & 1 & 1 \\
\texttt{vol\_nocorr} & raw Voronoi volume of the quasar ($V_{\rm raw}$)& 0.02 & 0.64 & 5.54 \\
\texttt{source\_id} & Gaia DR3 source identifier & - & - & - \\
\texttt{unwise\_objid} & unWISE DR1 source identifier & - & - & - \\
\texttt{sel\_func} & \emph{Quaia} selection function & 0.52 & 0.69 & 0.96 \\ 
\hline
\texttt{ID\_v} & void ID, if the quasar is a member & 55 & 16950 & 51439 \\
\texttt{ra\_v} & right ascension of void circumcentre & 0.03 & 172.72 & 359.99 \\
\texttt{dec\_v} & declination of void circumcentre & -81.69 & 1.11 & 83.84 \\
\texttt{z\_v} & redshift of void circumcentre & 0.80 & 1.46 & 2.20 \\
\texttt{ra\_v\_bc} & right ascension of void barycentre & 0.02 & 172.67 & 359.98 \\
\texttt{dec\_v\_bc} & declination of void barycentre & -81.56 & 1.38 & 83.99 \\
\texttt{z\_v\_bc} & redshift of void barycentre & 0.77 & 1.46 & 2.29 \\
\texttt{R\_eff\_mpc\_v} & effective radius of the parent void [$\mathrm{Mpc/h}$] & 43.11 & 135.93 & 266.91 \\
\texttt{delta\_min\_v} & minimum density fluctuation within the parent void & -0.76 & -0.56 & -0.01 \\
\texttt{delta\_avg\_v} & average density ($\delta_{\rm avg}$) fluctuation within the parent void & -0.32 & 0.00 & 0.94 \\
\texttt{lambda\_v} & $\delta_{\rm avg}\cdot R_{\rm eff}^{1.2}$ of the parent void (void type proxy) & -118.72 & 0.97 & 135.89 \\
\texttt{Theta\_eff\_v} & effective angular size of void [deg] & 0.89 & 2.68 & 5.52 \\
\texttt{EdgeFlag\_v} & edge flag of void (close to edge: 0, far from edge: 1) & 0 & 0 & 1 \\
\texttt{R\_over\_Rv} & relative distance of quasar to void circumcentre $(R/R_{\rm eff})$ & 0.01 & 1.33 & 4.18 \\ 
\hline
\texttt{ID\_c} & cluster ID, if the quasar is a member & 96 & 38026 & 94960 \\
\texttt{ra\_c} & right ascension of cluster circumcentre & 0.00 & 172.45 & 360.00 \\
\texttt{dec\_c} & declination of cluster circumcentre & -82.72 & 1.27 & 85.09 \\
\texttt{z\_c} & redshift of cluster circumcentre & 0.80 & 1.46 & 2.20 \\
\texttt{R\_eff\_mpc\_c} & $R_{\rm eff}$, effective radius of cluster & 45.06 & 86.28 & 157.26 \\
\texttt{delta\_max\_c} & maximum density fluctuation of a Voronoi cell in cluster & 0.04 & 1.79 & 49.80 \\
\texttt{delta\_avg\_c} &  average density fluctuation ($\delta_{\rm avg}$) of all Voronoi cells in cluster & -0.49 & 0.02 & 1.19 \\
\texttt{lambda\_c} & $\delta_{\rm avg}\cdot R_{\rm eff}^{1.6}$ of the parent cluster (cluster type proxy) & -656.00 & 21.74 & 1217.48 \\
\texttt{Theta\_eff\_c} & effective angular size of cluster [deg] & 0.82 & 1.70 & 3.64 \\
\texttt{EdgeFlag\_c} & edge flag of cluster (close to edge: 0, distant: 1) & 0 & 0 & 1 \\
\texttt{R\_over\_Rc} & relative distance of quasar to cluster circumcentre $(R/R_{\rm eff})$ & 0.00 & 1.11 & 5.26 \\

        \hline
    \end{tabular}
\tablefoot{The full table of 708,483 individual quasars is available at the CDS.}
\label{tab:value_added_table}
\end{table*}

\clearpage

\end{document}